\documentclass[journal=jpcafh,manuscript=article]{achemso}

\usepackage[normalem]{ulem}
\usepackage{amsmath,amssymb}
\usepackage{graphicx}
\usepackage{siunitx}
\usepackage{caption}
\usepackage{xcolor}
\usepackage[version=4]{mhchem}
\usepackage{chemfig}
\usepackage{subcaption}
\usepackage{tabularx}
\usepackage{array}
\usepackage[linktocpage,colorlinks=true,linkcolor=blue,urlcolor=blue,
citecolor=blue,hyperfigures]{hyperref}

\newcolumntype{Y}{>{\raggedright\arraybackslash}X}

\newcommand{\THz}{\ensuremath{\mathrm{THz}}}
\newcommand{\nre}{\ensuremath{n}}
\newcommand{\kim}{\ensuremath{\kappa}}
\newcommand{\cplx}{\ensuremath{\tilde n}}
\newcommand{\BCOp}{\ensuremath{\mathrm{BCO}^{+}}}

\title{Terahertz Time-Domain Spectroscopy and Density Functional Theory Analysis of Low-Frequency Vibrational Modes of a Benzoxazolium--Coumarin Donor--$\pi$--Acceptor Chromophore}

\author{Sidhanta Sahu}
\altaffiliation{These authors contributed equally to this work.}
\affiliation[IISERK]{Department of Physical Sciences, Indian Institute of Science Education and Research Kolkata, Mohanpur, West Bengal 741246, India.}

\author{Phalguna Krishna Das Vana}
\altaffiliation{These authors contributed equally to this work.}
\affiliation[IISERK]{Department of Physical Sciences, Indian Institute of Science Education and Research Kolkata, Mohanpur, West Bengal 741246, India.}

\author{Anupama Chauhan}
\affiliation[IISERK]{Department of Physical Sciences, Indian Institute of Science Education and Research Kolkata, Mohanpur, West Bengal 741246, India.}

\author{Poulami Ghosh}
\affiliation[IISERK]{Department of Physical Sciences, Indian Institute of Science Education and Research Kolkata, Mohanpur, West Bengal 741246, India.}

\author{Vijay Sai Krishna Cheerala}
\affiliation[SSSIHL]{Department of Chemistry, Sri Sathya Sai Institute of Higher Learning, Brindavan Campus, Kadugodi, Bangalore 560067, India.}

\author{Sanyam}
\affiliation[IITGN]{Department of Chemistry, Indian Institute of Technology Gandhinagar, Gandhinagar, Gujarat 382355, India.}

\author{C.~N. Sundaresan}
\affiliation[SSSIHL]{Department of Chemistry, Sri Sathya Sai Institute of Higher Learning, Brindavan Campus, Kadugodi, Bangalore 560067, India.}

\author{N. Kamaraju}
\affiliation[IISERK]{Department of Physical Sciences, Indian Institute of Science Education and Research Kolkata, Mohanpur, West Bengal 741246, India.}
\email{nkamaraju@iiserkol.ac.in}

\begin{document}

\begin{abstract}
To elucidate low-frequency vibrational modes, we investigate a benzoxazolium--coumarin (\BCOp{}) donor--\(\pi\)--acceptor derivative using transmission terahertz time-domain spectroscopy (THz--TDS). The retrieved complex refractive index reveals distinct modes at 0.62, 0.85, 1.30, 1.81, and 2.07~\THz. Gas-phase density functional theory (DFT) agrees well with these features and enables assignment of specific intramolecular motions. Together, THz--TDS and DFT identify the characteristic low-frequency modes of \BCOp{} and suggest their connection to intramolecular charge transfer-relevant nuclear motions, \textcolor{blue}{highlighting that THz--TDS can serve as} a sensitive probe of vibrational signatures in donor--\textcolor{blue}{$\pi$}--acceptor systems.
\end{abstract}

\section{Introduction}
Terahertz time-domain spectroscopy (THz--TDS) directly captures the subpicosecond electric-field pulse, enabling retrieval of both spectral amplitude and phase without relying on Kramers--Kronig analysis.\cite{neu_2018_jap, koch_2023_nrmp, ulbricht_2011_rmp} Low-frequency vibrational modes (below 3~\THz) serve as sensitive probes of molecular flexibility and weak noncovalent interactions, and can, when resonantly excited, influence charge-transfer pathways in organic molecules.\cite{chen_2021_saa,hou_2023_saa,paraipan_2023_applsci,ueno_2006_cheml,tu_2023_ijms,schweicher_2019_advmater,jacobs_2023_jpca} These collective motions encode conformational dynamics and intermolecular coupling in donor--\(\pi\)--acceptor (D--\(\pi\)--A) systems. Accordingly, THz--TDS provides access to this low-frequency regime and has been used to probe weak interactions,\cite{sun_2021_dalttrans} discriminate polymorphs,\cite{zhang_2016_jpcb} reveal phonons in organic semiconductor solids,\cite{schweicher_2019_advmater, koch_2023_nrmp, ulbricht_2011_rmp} and resolve H-bond network vibrations,\cite{paraipan_2023_applsci} \textcolor{blue}{as well as collective low-frequency motions in complex macromolecular systems and demonstrate sensitivity to hydration and conformational dynamics.\cite{walther_2003_chemphys}}

Many coumarin dyes are engineered as D--\(\pi\)--A chromophores---typically by placing a strong electron donor at the 6/7-position (e.g., dialkylamino/alkoxy) and coupling it through the coumarin \(\pi\)-system to an electron-accepting carbonyl or cationic heteroaromatic fragment---yielding analyte-responsive intramolecular charge transfer (ICT) and enabling optical and electrochemical sensing.\cite{khan_2023_aoc,kaushik_2016_acssensors,hara_2003_jpcb,wang_2007_advmater}
Benzoxazolium heteroaromatics provide a rigid, electron-accepting motif that supports planarity and extended conjugation, and are frequently used in ion-responsive fluorescent probes; recent benzoxazole-based sensors, for example, enable cascade recognition of CN\(^-\) and Fe\(^{3+}\).\cite{cheerala_2023_langmuir,tohora_2023_jphotochemA,yang_2017_analmeth} Fusing these motifs yields benzoxazolium--coumarin (\BCOp{}) derivatives, prototypical D--\(\pi\)--A molecules with extended \(\pi\)-conjugation, pronounced ICT, and conformationally flexible scaffolds relevant to light--matter interactions and nonlinear optics.\cite{baseia_2017_crystals,zhao_1990_chemmater,shang_1998_josab} Notably, this scaffold has been deployed as a cyanide sensor, highlighting its ICT-based reactivity and analyte sensitivity.\cite{palanisamy_2024_jfluoresc}

\textcolor{blue}{Here we study a structurally simple but functionally rich D--$\pi$--A chromophore. \BCOp{} is small enough that individual low-frequency modes can be resolved and assigned, yet it retains the essential features of ICT-active D--$\pi$--A systems. Our strategy is therefore to use \BCOp{} as a model platform where selected vibrational modes can be connected, one by one, to well-defined structural coordinates along the ICT pathway.}

Despite extensive optical studies, systematic investigations of the low-energy IR-active modes of \BCOp{} in the THz region remain limited. Motivated by the sensitivity of \BCOp{} to the local environment and modulation of ICT pathways, we employ THz--TDS in transmission to understand its low-frequency vibrational dynamics. The complex refractive index retrieved over 0.43--2.51~\THz reveals distinct resonances at 0.62, 0.85, 1.30, 1.81, and 2.07~\THz. Complementary gas-phase DFT reproduces these features and assigns them to specific intramolecular motions involving torsion and bridge deformations within the D--\(\pi\)--A framework. Together, THz--TDS and DFT establish the characteristic low-frequency modes of \BCOp{} and suggest their relevance to \textcolor{blue}{ICT-related nuclear motions}, highlighting the potential of THz--TDS as a sensitive probe of intramolecular interactions and analyte-responsive dynamics in D--\(\pi\)--A systems.

\section{Experimental Methods}
\subsection{Materials}
The chemicals required for the synthesis and spectroscopic studies were purchased from Sigma-Aldrich, TCI, and Alfa Aesar and were used as received. The progress of the reactions was monitored by thin-layer chromatography (TLC) on Merck silica gel 60 F\(_{254}\) plates. \({}^{1}\)H and \({}^{13}\)C NMR spectra were recorded on a Bruker 500\,MHz FT-NMR spectrometer using tetramethylsilane (TMS) as an internal reference and dimethyl sulfoxide-\(d_6\) (DMSO-\(d_6\)) as the solvent. UV--Vis absorption spectra were recorded on a LAMBDA 365 UV--Vis spectrophotometer (PerkinElmer, Waltham, MA, USA) using a quartz cuvette. Fluorescence spectra were recorded on a Fluorolog-QM spectrofluorometer (HORIBA Scientific, Piscataway, NJ, USA).

\subsection{Synthesis}
\begin{scheme}[htbp]
  \centering
  \includegraphics[width=\linewidth]{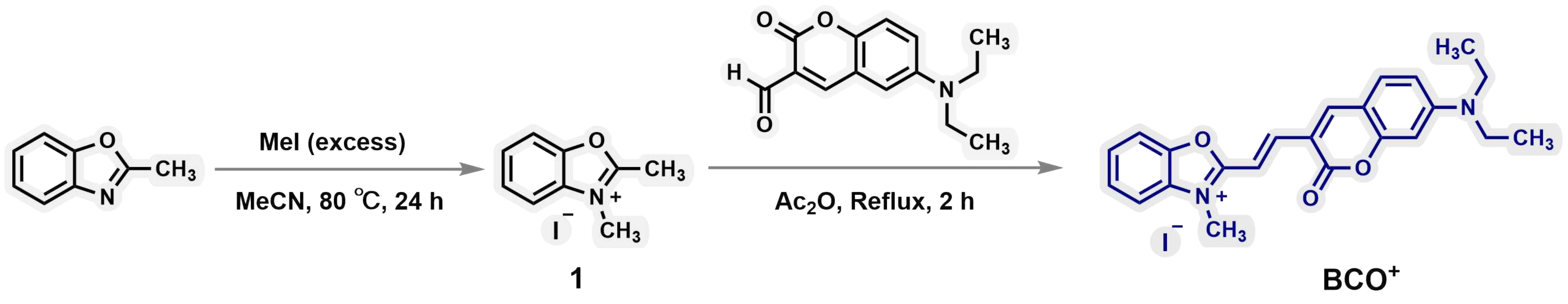}
  \caption{Synthetic route for the preparation of the benzoxazolium--coumarin derivative (\BCOp{}) from 2-methylbenzoxazole.}
  \label{sch:synthesis}
\end{scheme}

\subsubsection*{\textit{2,3-dimethylbenzo[d]oxazol-3-ium iodide (1).}}
2-methylbenzoxazole (0.025 mol) and methyl iodide (0.125 mol) were mixed in 15 mL of acetonitrile and heated to reflux for 24 hours. After the reaction, ethyl acetate was added to the reaction mixture to induce precipitation. The precipitate was filtered, washed with ethyl acetate, and dried in a vacuum oven. Yield: 81\%; \({}^{1}\)H NMR (400 MHz, DMSO-\(d_6\)) \(\delta\) 8.13 (m, 2H), 7.85 (m, 2H), 4.09 (s, 3H), 3.01 (s, 3H); \({}^{13}\)C NMR (100 MHz, DMSO-\(d_6\)) \(\delta\) 169.13, 147.31, 130.42, 128.76, 127.80, 114.55, 112.96, 32.77, 13.52.

\subsubsection*{\textit{(E)-2-(2-(7-(diethylamino)-2-oxo-2H-chromen-3-yl)vinyl)-3-methylbenzo[d]oxazol-3-ium iodide (BCO\textsuperscript{+}).}}
Benzoxazolium salt (1) (0.00145 mol) and the coumarin aldehyde (0.00147 mol) were mixed in 20 mL of acetic anhydride and heated under reflux for 2 hours. The mixture was allowed to cool to room temperature. The precipitated solid was collected by filtration, washed with 20 mL of ethyl acetate, and dried. Yield: 79\%; \({}^{1}\)H NMR (400 MHz, DMSO-\(d_6\)) \(\delta\) 8.63 (s, 1H), 8.19 (d, \(J = 15.4\) Hz, 1H), 8.05 (m, 2H), 7.72 (m, 2H), 7.69 (d, \(J = 15.5\) Hz, 1H), 7.57 (d, \(J = 9.1\) Hz, 1H), 6.88 (dd, \(J = 9.1, 2.4\) Hz, 1H), 6.67 (d, \(J = 2.4\) Hz, 1H), 4.08 (s, 3H), 3.54 (q, \(J = 7.1\) Hz, 4H), 1.17 (t, \(J = 7.1\) Hz, 6H); \({}^{13}\)C NMR (100 MHz, DMSO-\(d_6\)) \(\delta\) 162.67, 159.64, 157.47, 153.61, 148.96, 147.07, 145.74, 132.12, 131.37, 128.46, 127.50, 114.01, 112.31, 111.49, 111.06, 109.00, 101.94, 96.53, 44.80, 32.04, 12.49 (see the SI, Figs.~S1--S2 for $^1$H/$^{13}$C NMR spectra).

\subsection{THz--TDS Setup}
\begin{figure}[htbp]
  \centering
  \includegraphics[width=0.85\linewidth]{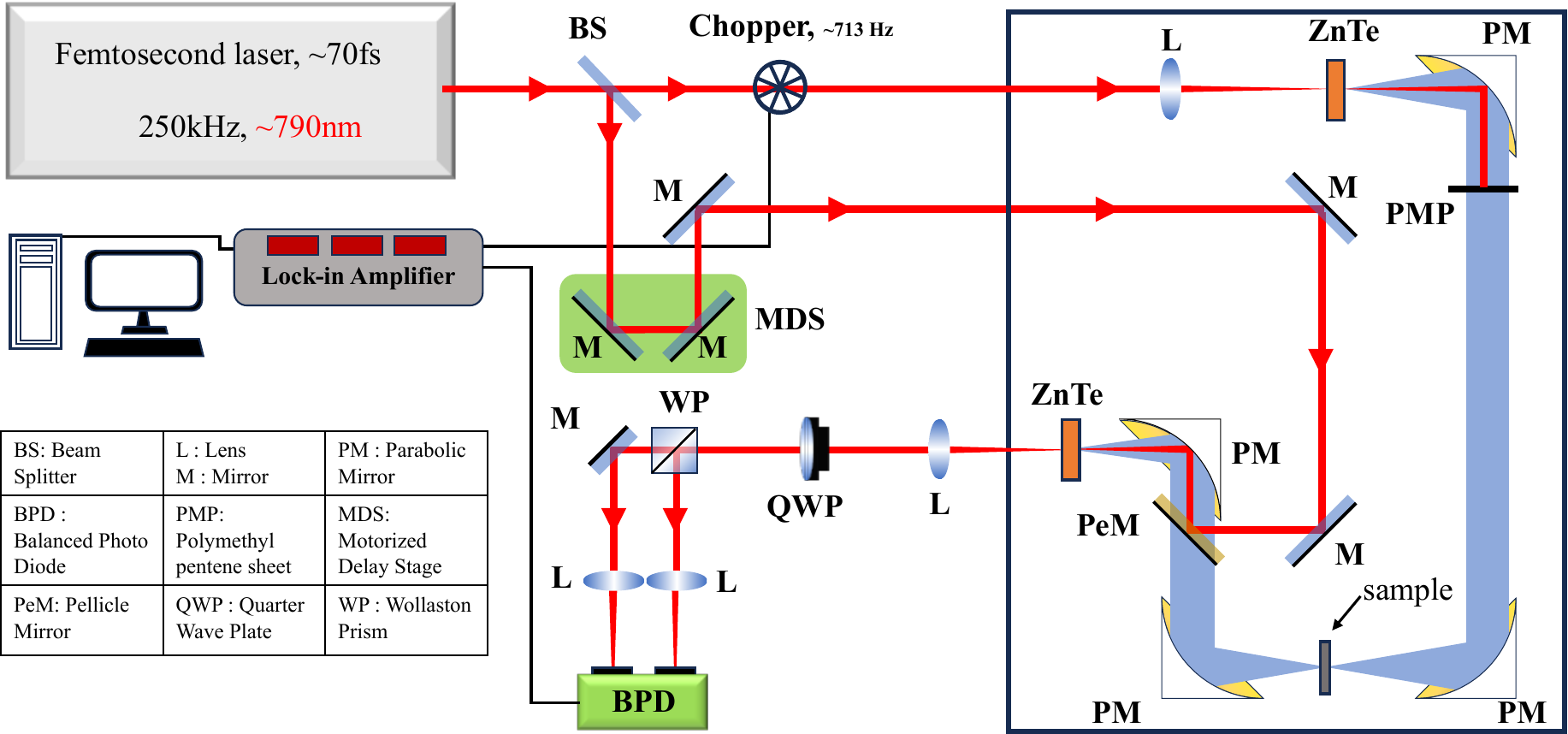}
  \caption{Schematic of the home-built THz--TDS setup in transmission geometry.}
  \label{fig:setup}
\end{figure}

THz--TDS measurements were performed in transmission geometry. Femtosecond pulses at 790~nm (\(\sim\)70~fs, \(\sim\)250~kHz) were supplied by a Ti:sapphire regenerative amplifier (Coherent RegA~9050). THz pulses were generated and detected by a \(\langle 110\rangle\)-cut ZnTe crystal. The THz detection unit consists of (i) a quarter-wave plate, (ii) a Wollaston prism, and (iii) a home-built balanced photodiode (see Fig.~\ref{fig:setup}). The setup was purged with dry N\(_2\) to reduce H\(_2\)O absorption. The \BCOp{} sample was homogenized and pressed into a pellet with thickness \(d=\SI{459\pm7}{\micro\meter}\) (see the SI, Table~S1). \textcolor{blue}{Multiple micrometer readings across the pellet surface yield only a few-percent spread around this value (SI, Table~S1). Repeatability of the time-domain traces is shown in SI, Figs.~S5--S6, where five independent runs for both the air reference and the \BCOp{} pellet are overlaid; run-to-run variations are small compared to the signal amplitude. Taken together, these measurements indicate that thickness variations and temporal fluctuations are minor relative to the spectral features discussed below.} The complex refractive index was retrieved using a standard Fresnel transmission model.\cite{duvillaret_1996_jstqe,mukherjee_2021_epjst}

\subsection{Computational Methods}
\label{sec:compdetails}
All quantum-chemical calculations were performed with \textsc{Gaussian}~16, Rev.~B.01.\cite{frisch_2016_gaussian16} Structures were optimized at the B3LYP-D3/6-311++G(d,p) level with Grimme’s D3 dispersion correction.\cite{becke_1993_jcp,lee_1988_prb,grimme_2010_jcp} An ultrafine integration grid and tight SCF thresholds were used throughout; geometry optimizations employed very-tight criteria (Opt=VeryTight) with analytical Hessians at the first step (CalcFC). \textcolor{blue}{For isolated organic chromophores, hybrid density functionals such as B3LYP combined with triple-$\zeta$ Pople basis sets including polarization and diffuse functions (e.g.,
6-311++G(d,p)) are widely used and benchmarked for vibrational frequency calculations and spectroscopic modeling, including in $\pi$-conjugated chromophores and related molecular
systems.\cite{andersson_2005_jpca,greenwood_2014_jpcl,tang_2023_jpcl} Vibrational frequency analyses confirmed that all optimized geometries correspond to true minima on the potential-energy surface, as evidenced by the absence of imaginary frequencies.} To compare with experiment, a single uniform frequency-scaling factor was applied to the DFT values (reported with the assignments), consistent with established practice for vibrational frequency benchmarking.\cite{scott_1996_jpc,merrick_2007_jpca,alecu_2010_jctc}

\section{Results and Discussion}
\begin{figure}[htbp]
  \centering
  \begin{subfigure}[b]{0.45\linewidth}
    \centering
    \includegraphics[width=\linewidth]{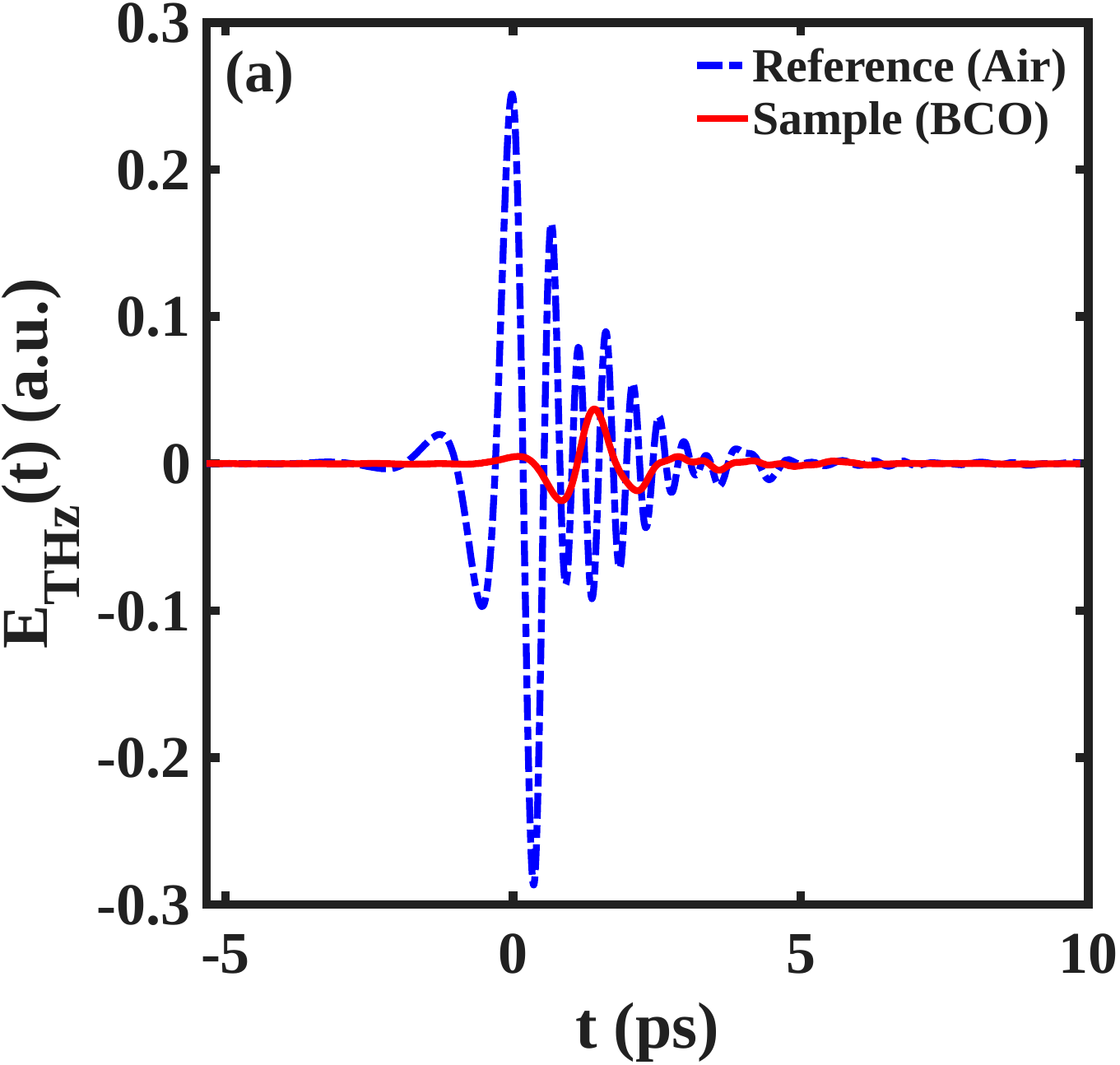}
    \phantomsubcaption\label{fig:tds_fft_a}
  \end{subfigure}\hfill
  \begin{subfigure}[b]{0.45\linewidth}
    \centering
    \includegraphics[width=\linewidth]{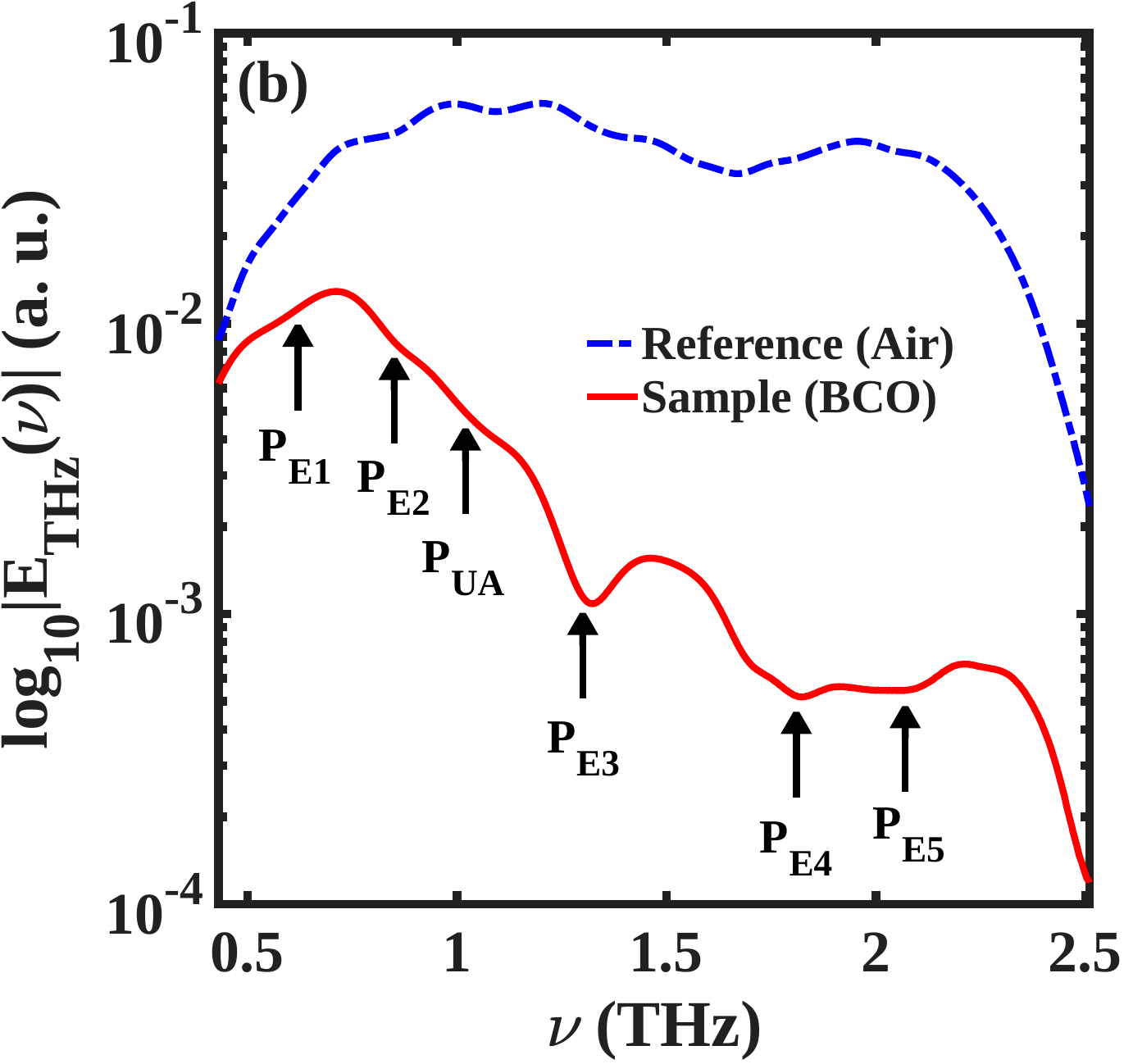}
    \phantomsubcaption\label{fig:tds_fft_b}
  \end{subfigure}
  \begin{subfigure}[b]{0.45\linewidth}
    \centering
    \includegraphics[width=\linewidth]{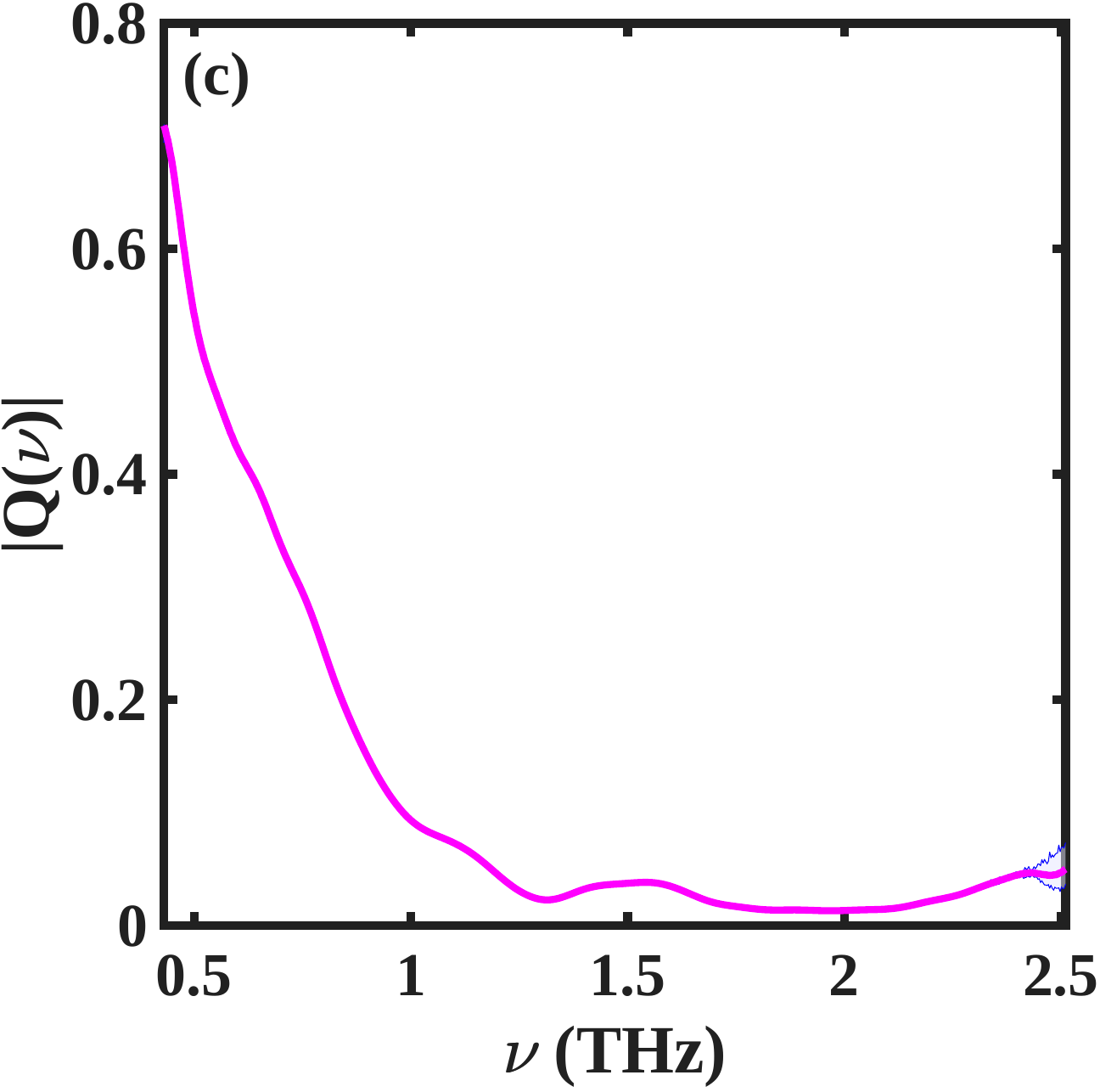}
    \phantomsubcaption\label{fig:tds_fft_c}
  \end{subfigure}\hfill
  \begin{subfigure}[b]{0.45\linewidth}
    \centering
    \includegraphics[width=\linewidth]{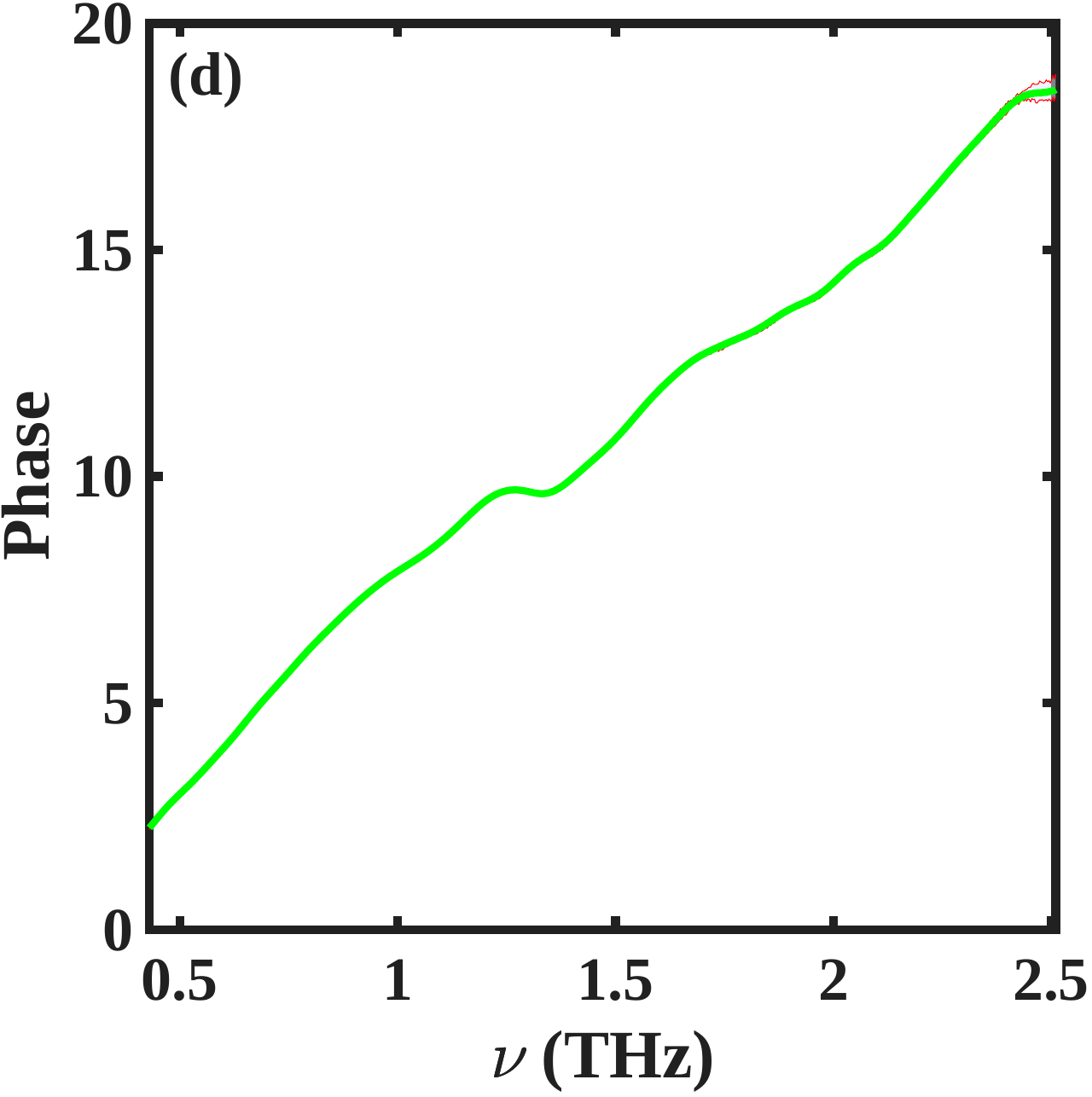}
    \phantomsubcaption\label{fig:tds_fft_d}
  \end{subfigure}
\caption{THz--TDS overview for the \BCOp{} pellet: (a) time-domain waveforms reference (blue dotted) vs.\ sample (red solid), (b) FFT spectra reference (blue dotted) vs.\ sample (red solid) defining the 0.43--2.51~\THz analysis window, (c) magnitude of the complex transmission \(|Q(\nu)|\), where the thick solid magenta curve shows the mean and the thin blue curves indicate the upper and lower uncertainty bounds at each frequency, and (d) unwrapped phase of \(Q(\nu)\) with the thick green curve as the mean phase and thin red curves as the corresponding phase uncertainty bounds used to retrieve the complex refractive index.}

  \label{fig:tds_fft}
\end{figure}

Figure~\ref{fig:tds_fft} summarizes the experimental THz--TDS measurements. The time-domain traces in Fig.~\ref{fig:tds_fft_a} show reduced peak amplitude and a clear temporal delay for the \BCOp{} pellet relative to the reference, consistent with absorption and an effective refractive index near \(\sim\)1.8. The corresponding Fast Fourier Transform (FFT) spectra in Fig.~\ref{fig:tds_fft_b} define a usable bandwidth of 0.43--2.51~\THz, within which the main absorption features of \BCOp{} are evident and subsequently analyzed. \textcolor{blue}{The total time window of the recorded THz waveform is
\( \approx 15~\text{ps}\), corresponding to a nominal frequency
resolution of \(\Delta \nu \approx 0.066~\text{THz}\)
(\(\approx 66~\text{GHz}\)), which sets the spectral resolution of
the present THz--TDS measurements. Zero-padding was used only to interpolate the FFT grid and does not improve the true spectral resolution.
}

We compute the complex transmission \(Q_{\mathrm{exp}}(\nu)\equiv E_{\mathrm{sam}}(\nu)/E_{\mathrm{ref}}(\nu)\) (Figs.~\ref{fig:tds_fft_c}--\ref{fig:tds_fft_d}) and determine the complex refractive index by minimizing the error function \(\zeta(\nu)\) in Eq.~\ref{eq:zeta}, \textcolor{blue}{where \(\nu\) is reported in \THz, \(d\) is the thickness of the pellet, and \(c\) is the speed of light in vacuum.}

\begin{equation}
\zeta(\nu)=
\left\lvert
\frac{4\,\cplx(\nu)}{\bigl(1+\cplx(\nu)\bigr)^{2}}
\frac{\mathrm{e}^{\,i\frac{2\pi \nu d}{c}\bigl(\cplx(\nu)-1\bigr)}}
     {1-\left(\frac{1-\cplx(\nu)}{1+\cplx(\nu)}\right)^{\!2}
        \mathrm{e}^{\,i\frac{4\pi \nu d}{c}\cplx(\nu)}}
- Q_{\mathrm{exp}}(\nu)
\right\rvert^{2}.
\label{eq:zeta}
\end{equation}

\begin{figure}[htbp]
  \centering
  \begin{subfigure}[b]{0.48\linewidth}
    \centering
    \includegraphics[width=\linewidth]{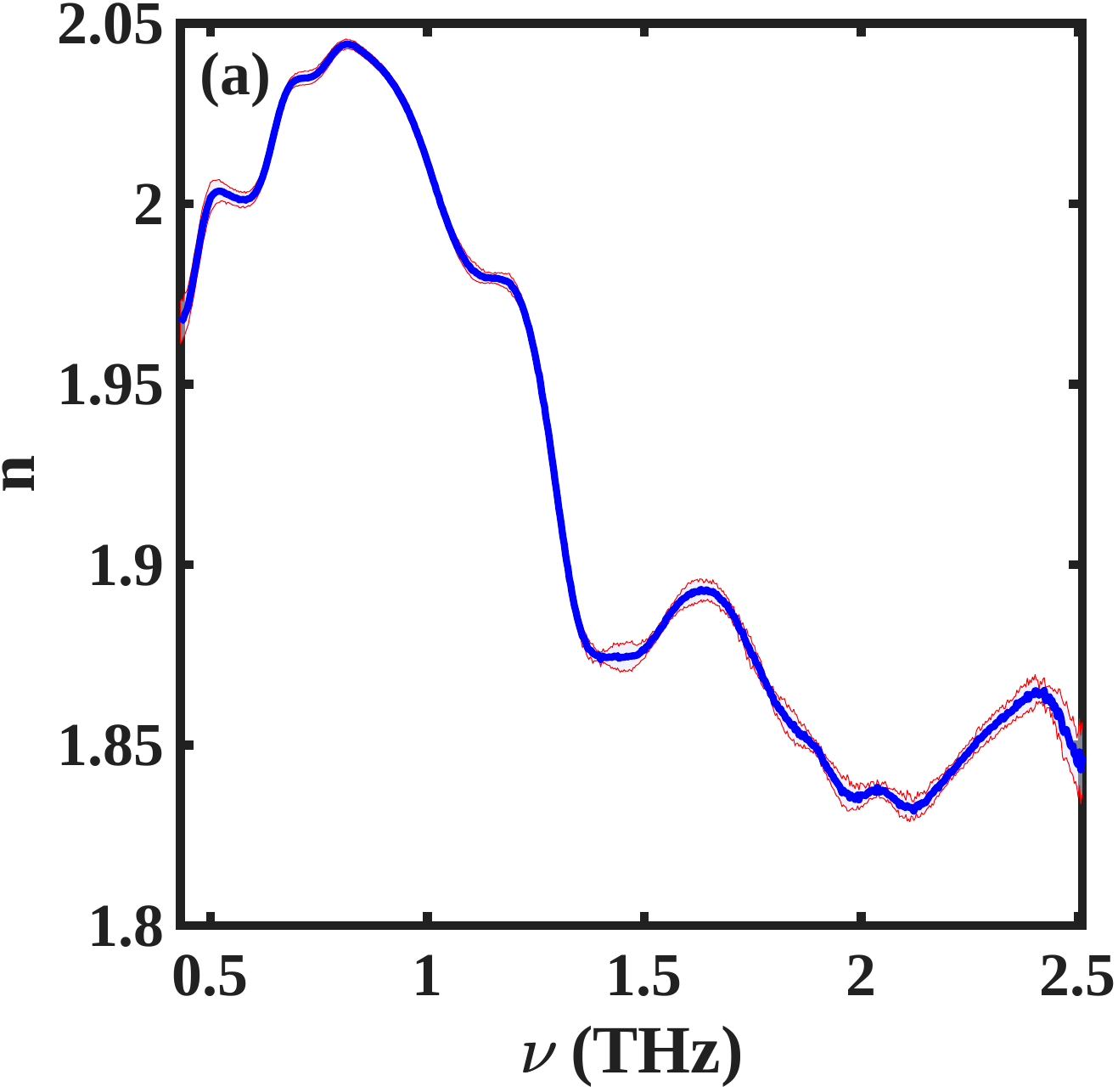}
    \phantomsubcaption\label{fig:optconst_a}
  \end{subfigure}\hfill
  \begin{subfigure}[b]{0.48\linewidth}
    \centering
    \includegraphics[width=\linewidth]{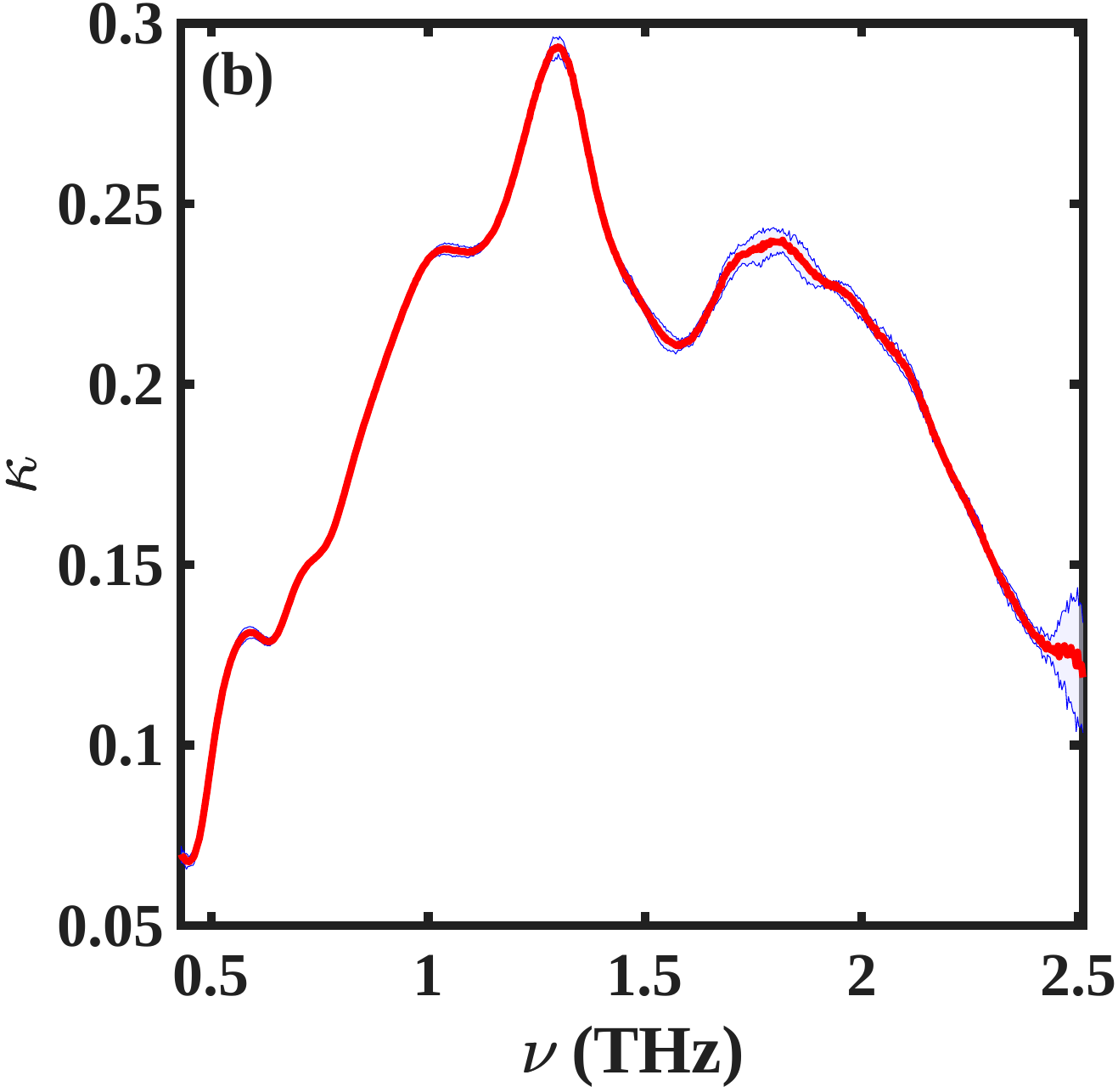}
    \phantomsubcaption\label{fig:optconst_b}
  \end{subfigure}
\caption{Retrieved complex refractive index for \BCOp{} from THz--TDS: (a) real refractive index \(\nre(\nu)\), plotted as a thick solid blue curve for the mean spectrum with thin red curves indicating the upper and lower uncertainty bounds at each frequency, and (b) extinction coefficient \(\kim(\nu)\), plotted as a thick solid red curve for the mean spectrum with thin blue curves indicating the corresponding uncertainty bounds at each frequency, together forming error bands for both optical constants.}

  \label{fig:optconst}
\end{figure}

This procedure yields the complex refractive index \(\cplx(\nu)=\nre(\nu)+i\kim(\nu)\), shown in Fig.~\ref{fig:optconst}. The real part \(\nre(\nu)\) (Fig.~\ref{fig:optconst_a}) exhibits normal dispersion with changes near absorptive regions. The extinction coefficient \(\kim(\nu)\) (Fig.~\ref{fig:optconst_b}) reveals five principal absorption regions centered near 0.62, 0.85, 1.30, 1.81, and 2.07~\THz, and a weaker line at 1.02~\THz.

\begin{figure}[htbp]
  \centering
  \includegraphics[width=0.5\linewidth]{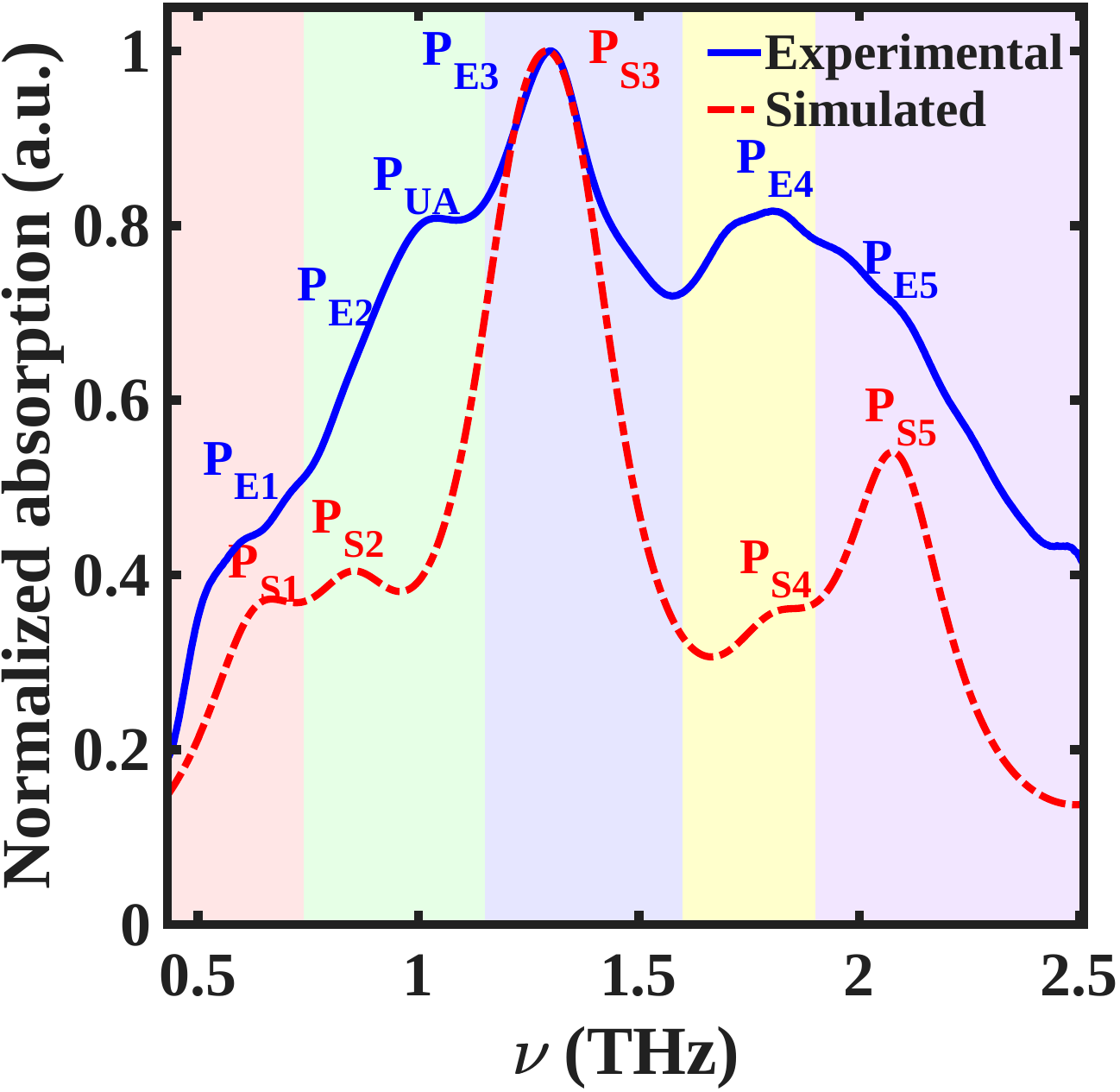}
  \caption{Experimental extinction coefficient \(\kim(\nu)\) (blue solid) overlaid with the scaled DFT spectrum (red dashed).}
  \label{fig:overlay}
\end{figure}

To understand the nature of these vibrational modes, we performed a gas-phase DFT analysis of \BCOp{} at the B3LYP-D3/6-311++G(d,p) level. Normal-mode frequencies and IR intensities were obtained at a fully optimized minimum (no imaginary frequencies). To compare with experiment, we mapped each calculated frequency \(\nu_{\text{calc}}\) to the observed value via a single least-squares scale factor, \(\nu_{\text{obs}}=s\,\nu_{\text{calc}}\), which yielded \(s=1.24\) with \(\mathrm{RMSE}=0.0145\)~\THz across the 0.43--2.51~\THz window (see the SI, Section S3; Table S2). After scaling, the principal simulated positions are \(P_{S1}=0.624\)~\THz, \(P_{S2}=0.843\)~\THz, the \(P_{S3}\) doublet at \(\{1.219,\,1.336\}\)~\THz, \(P_{S4}=1.790\)~\THz, and \(P_{S5}=2.066\)~\THz, in good agreement with the experimentally observed modes. Inspection of the corresponding eigenvectors indicates inter-ring torsion/libration and coupled skeletal/bridge deformations that act along the D--\(\pi\)--A axis and therefore carry sizable IR intensity in the THz region. \textcolor{blue}{In Fig.~\ref{fig:overlay}, we therefore plot the experimental extinction coefficient together with the scaled DFT spectrum (see Table~\ref{tab:assign} for the comparison of the various modes, \(P_{S1}\)--\(P_{S5}\), along with the calculated relative intensities).}

\begin{table}[t]
  \centering
  \setlength{\tabcolsep}{6pt}
  \begin{tabularx}{\linewidth}{l c c Y}
    \hline
    \textbf{Mode} & \textbf{Frequency (\THz)} & \textcolor{blue}{\textbf{Rel. Intensity}} & \textbf{Mode Assignment} \\
    \hline
    $P_{S1}$ & 0.62 & \textcolor{blue}{0.42} & Inter-ring torsion/libration about the D--$\pi$--A bridge \\
    $P_{S2}$ & 0.84 & \textcolor{blue}{0.36} & Ring libration with minor bridge participation \\
    $P_{S3}$ & 1.22 & \textcolor{blue}{0.93} & Skeletal deformation + bridge twist (coupled; larger amplitude on the coumarin ring) \\
             & 1.34 & \textcolor{blue}{1.00} & Bridge twist + benzoxazolium wag (coupled) \\
    $P_{S4}$ & 1.79 & \textcolor{blue}{0.27} & Localized deformation (bridge shear/ring) \\
    $P_{S5}$ & 2.07 & \textcolor{blue}{0.80} & Higher-frequency coupled ring deformation \\
    \hline
    \multicolumn{4}{l}{\footnotesize $P_{S3}$ is a doublet (1.22 and 1.34~\THz) that merges into a single peak. See the SI Movies S1--S6 for animations.}\\
    \hline
  \end{tabularx}
  \caption{Low-frequency IR-active modes of BCO\textsuperscript{+} predicted by DFT.}
  \label{tab:assign}
\end{table}

Table~\ref{tab:assign} summarizes the IR-active modes of \BCOp{} resolved in this work (see the SI S4-S5 for snapshots and animations). The lowest-frequency mode \(P_{E1}\) at 0.62~\THz is reproduced by \(P_{S1}\) and corresponds to an inter-ring torsion/libration about the \BCOp{} bridge. The second mode \(P_{E2}\) at 0.85~\THz is matched by \(P_{S2}\) and is best described as a ring libration with minor bridge participation. The intense mode \(P_{E3}\) at 1.30~\THz is dominated by two intramolecular contributors (\(P_{S3}\)), whose coupled skeletal/bridge character \textcolor{blue}{is expected to influence} the ICT axis. At higher frequencies, \(P_{E4}\) at 1.81~\THz is due to localized deformation (\(P_{S4}\)), and \(P_{E5}\) at 2.07~\THz is due to higher-frequency coupled ring deformation (\(P_{S5}\)). The near-uniform blue shift (single factor \(s=1.24\)) indicates systematic hardening of these soft modes in the solid state relative to the gas phase; we attribute this to intermolecular constraints in the pellet---packing, local electrostatics, and weak contacts---that stiffen torsional and librational coordinates. The small residual RMSE implies that environment-induced shifts are broadly mode-independent at this resolution, while the unassigned \(P_{UA}=1.02\)~\THz feature is \textcolor{blue}{plausibly due to condensed-phase activation of a nominally weak or IR-inactive mode, for example via packing-induced symmetry breaking or weak intermolecular coupling, without a clear one-to-one gas-phase counterpart.\cite{takahashi_2014_crystals,hadjiivanov_2021_chemrev}}

\begin{figure}[htbp]
  \centering
  \begin{subfigure}[b]{0.48\linewidth}
    \centering
    \includegraphics[width=\linewidth]{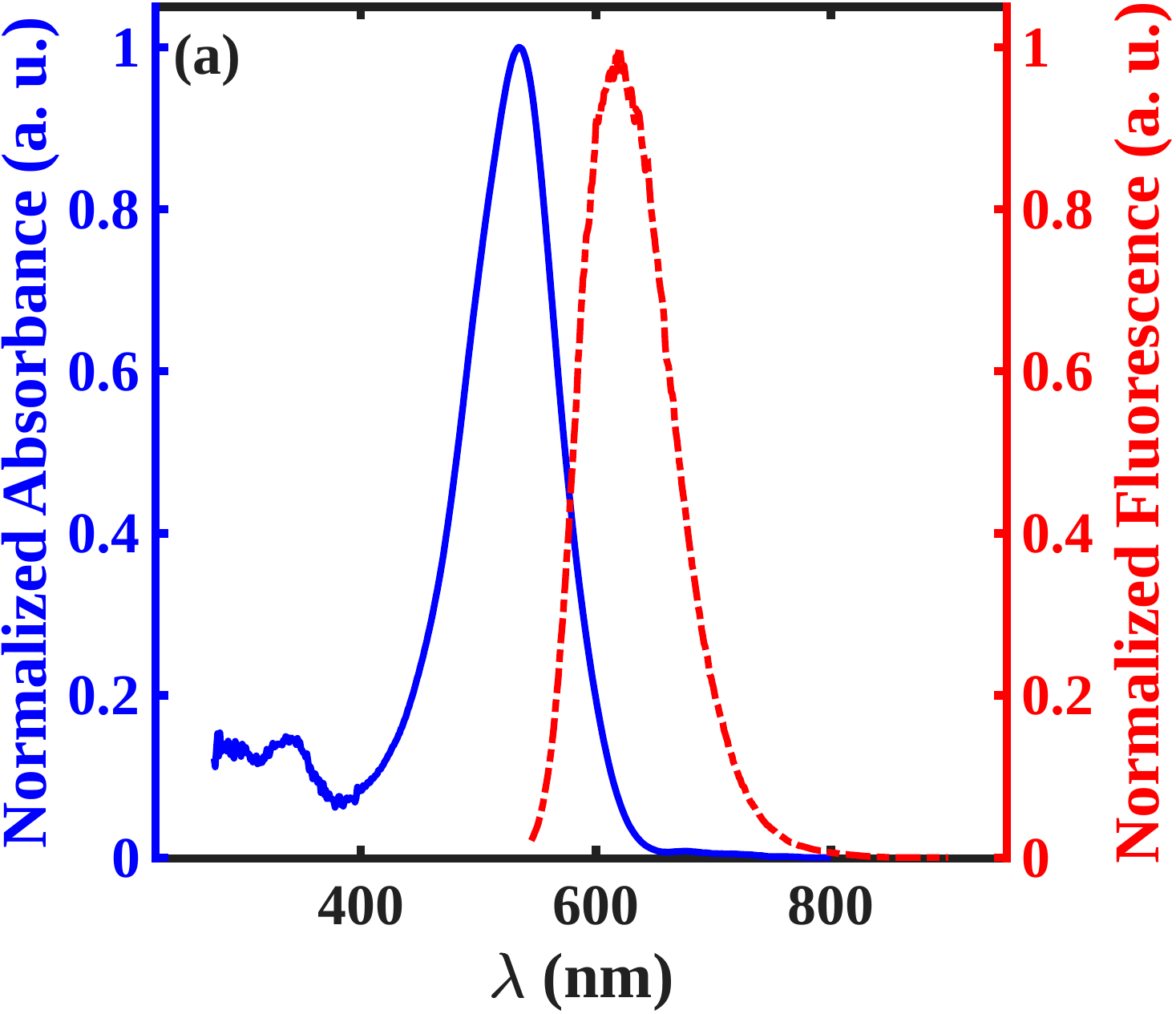}
    \phantomsubcaption\label{fig:uv_vis_fl}
  \end{subfigure}\hfill
  \begin{subfigure}[b]{0.48\linewidth}
    \centering
    \includegraphics[width=\linewidth]{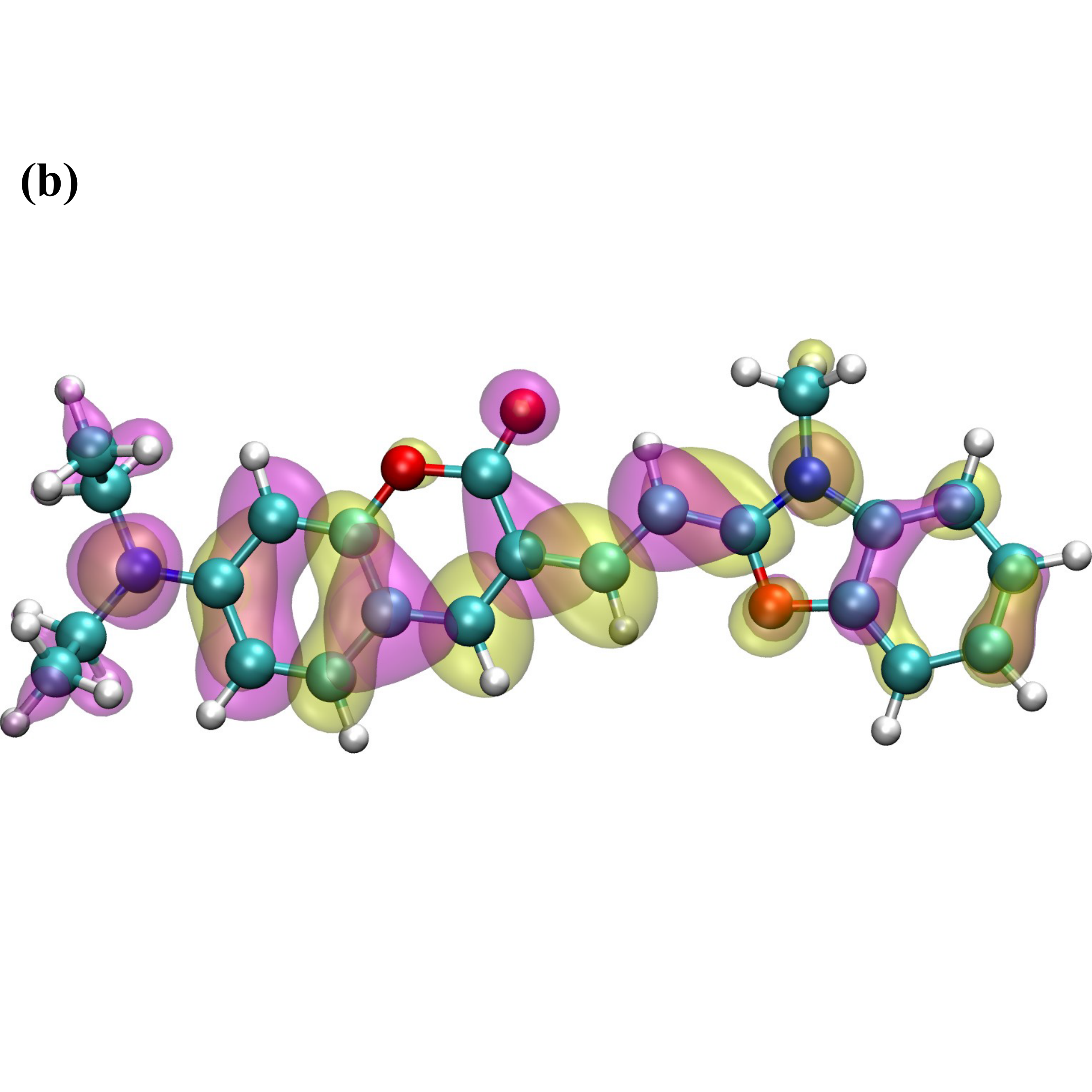}
    \phantomsubcaption\label{fig:HOMO_LUMO}
  \end{subfigure}
  \caption{(a) UV--Vis absorption (blue solid) and steady-state fluorescence (\(\lambda_{\mathrm{ex}}=530\,\mathrm{nm}\); red dashed) spectra of \BCOp{} in DMSO at 298\,K. (b) HOMO--LUMO isosurfaces (isovalue \(0.02\)); the HOMO is localized on the donor with partial backbone delocalization, and the LUMO on the acceptor, consistent with D--\(\pi\)--A ICT.}
  \label{fig:uv_homo_panel}
\end{figure}

The \BCOp{} derivative represents a highly polarized D--\(\pi\)--A molecular architecture that facilitates ICT through an extended conjugated framework. The N,N-diethylamino group at the 7-position of the coumarin ring acts as a strong electron donor, while the intrinsic carbonyl functionality of coumarin and the cationic benzoxazolium fragment serve as potent electron acceptors. The trans-vinylene linkage connecting these moieties provides a delocalized \(\pi\)-bridge that mediates efficient electronic communication between the donor and acceptor termini. Such push--pull structures typically display large Stokes shifts because the relaxed \(S_1\) state has pronounced ICT character; consistent with Fig.~\ref{fig:uv_vis_fl}, \BCOp{} in DMSO at 298~K shows \(\lambda_{\mathrm{abs}}^{\max}=\SI{535}{\nano\meter}\) and \(\lambda_{\mathrm{em}}^{\max}=\SI{620}{\nano\meter}\), yielding \(\Delta\lambda=\SI{85}{\nano\meter}\) (\(\Delta E=\SI{0.318}{\electronvolt}\)). Additional steady-state spectra are provided in the SI (Figs.~S3--S4). The broad, featureless emission band is typical of relaxed ICT states in polar media.\cite{satpati_2009_photochemphotobiol,liu_2022_chemosensors} The HOMO/LUMO topology (Fig.~\ref{fig:HOMO_LUMO})---HOMO on the coumarin donor and LUMO on the benzoxazolium acceptor---corroborates this assignment. Mechanistically, the torsional/bridge modes resolved by THz--TDS (\(P_{S1}\)--\(P_{S3}\)) \textcolor{blue}{when excited resonantly with an intense THz pulse, are expected to transiently alter D--\(\pi\)--A planarity and orbital overlap, and thus to influence the ICT gap}, rationalizing the large Stokes shift observed in Fig.~\ref{fig:uv_vis_fl}.

\section{Conclusion}
THz--TDS reveals a set of low-energy, IR-active modes in the benzoxazolium--coumarin (\BCOp{}) D--\(\pi\)--A scaffold, with five prominent peaks at \(P_{E1}=0.62\), \(P_{E2}=0.85\), \(P_{E3}=1.30\), \(P_{E4}=1.81\), and \(P_{E5}=2.07\)~\THz, along with a weaker unassigned feature at \(P_{UA}=1.02\)~\THz. Gas-phase DFT IR mode calculations agree well with experiment and enable the assignment of these observed modes: The \(P_{E3}\) peak arises from a merged doublet within \(P_{S3}\), while \(P_{E1}\) and \(P_{E2}\) originate from torsion/libration. The presence of \(P_{UA}\) without a gas-phase counterpart highlights condensed-phase contributions beyond single-molecule theory \textcolor{blue}{such as symmetry breaking and intermolecular constraints}. Selective THz excitation of \(P_{S1}\)--\(P_{S3}\) could enable direct tests of vibronic gating of ICT, offering a route to control charge-transfer pathways through targeted low-frequency mode excitation. \textcolor{blue}{More broadly, combining THz--TDS with DFT mode analysis in this well-defined D--$\pi$--A chromophore provides a model-system framework that can be extended to families of related chromophores and to time-resolved THz experiments that directly probe vibronically mediated charge transfer.}

\section*{Acknowledgments}
The authors thank the Ministry of Education (MoE), Government of India, for financial support, and IISER Kolkata for providing the infrastructural facilities to carry out this research. S.S., A.C., and P.G. acknowledge the University Grants Commission (UGC), IISER Kolkata, the Council of Scientific and Industrial Research (CSIR), and the Department of Science and Technology (DST)--Innovation in Science Pursuit for Inspired Research (INSPIRE) for their research fellowships. The authors thank Prof. Venkataramanan Mahalingam and Mr. Deepanjan Patra for assistance with fluorescence measurements. The authors thank Prof. Ratheesh K Vijayaraghavan and Ms. Swati Chakraborty for assistance with NMR measurements. The authors thank Pedisetti Venkatesh for assistance with the UV--Visible measurements.

\bibliography{references}

\end{document}